\documentclass[lettersize, journal]{IEEEtran}

\usepackage{cite}
\usepackage{amsmath,amssymb,amsfonts}
\usepackage{algorithm}
\usepackage{algpseudocode}
\usepackage{graphicx}
\usepackage{textcomp}
\usepackage{xcolor}
\usepackage{listings}
\usepackage{makecell}
\usepackage{pifont}
\usepackage[hidelinks]{hyperref} 
\usepackage[caption=false,font=normalsize,labelfont=sf,textfont=sf]{subfig}

\usepackage{multirow}
\usepackage{array}      % To modify table rule thickness

\newcommand{\orcidd}[1]{\href{https://orcid.org/#1}{\includegraphics[scale=0.18]{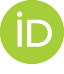}}\hspace{0mm}}

\usepackage{eso-pic}
\usepackage{xcolor}
\usepackage{graphicx}

\AddToShipoutPictureBG*{%
  \AtPageUpperLeft{%
    \put(80,-12){\parbox{\textwidth}{\centering\fontsize{7}{8}\selectfont\textcolor{black}{This article has been accepted for publication in IEEE Transactions on Computational Social Systems. This is the author's version which has not been fully\\edited and content may change prior to final publication. Citation information: DOI 10.1109/TCSS.2025.3565377}}}%
  }
  \AtPageLowerLeft{%
    \put(80,10){\parbox{\textwidth}{\centering\fontsize{7}{9}\selectfont\textcolor{black}{© 2025 IEEE. All rights reserved, including rights for text and data mining and training of artificial intelligence and similar technologies. Personal use is permitted,\\but republication/redistribution requires IEEE permission. See \href{https://www.ieee.org/publications/rights/index.html}{https://www.ieee.org/publications/rights/index.html} for more information.}}}%
  }
}

\begin{document}

\title{Crowd: A Social Network Simulation Framework\\}

\author{Ann Nedime Nese~Rende\orcidd{0009-0006-3561-9710},
        Tolga~Yilmaz\orcidd{0000-0001-8617-9301},
        and~Özgür~Ulusoy\orcidd{0000-0002-6887-3778},~\IEEEmembership{Member,~IEEE}% <-this % stops a space
\thanks{Manuscript received -. \textit{(Corresponding author: Ann Nedime Nese~Rende)}}
\thanks{Ann Nedime Nese Rende and Özgür Ulusoy are with the Department of Computer Engineering, Bilkent
University, 06800 Ankara, Türkiye (e-mail: nedime.rende@bilkent.edu.tr, oulusoy@cs.bilkent.edu.tr).}% <-this % stops a space
\thanks{Tolga Yilmaz is with the Department of Computer Science, University of Oxford, Oxford, United Kingdom (e-mail: tolga.yilmaz@cs.ox.ac.uk).}% <-this % stops a space

}

\markboth{IEEE TRANSACTIONS ON COMPUTATIONAL SOCIAL SYSTEMS}%
{Rende \MakeLowercase{\textit{et al.}}: Crowd: A Social Network Simulation Framework}


\maketitle

\begin{abstract}
To observe how individual behavior shapes a larger community's actions, agent-based modeling and simulation (ABMS) has been widely adopted by researchers in social sciences, economics, and epidemiology. While simulations can be run on general-purpose ABMS frameworks, these tools are not specifically designed for social networks and, therefore, provide limited features, increasing the effort required for complex simulations. In this paper, we introduce \textit{Crowd}, a social network simulator that adopts the agent-based modeling methodology to model real-world phenomena within a network environment. Designed to facilitate easy and quick modeling, Crowd supports simulation setup through YAML configuration and enables further customization with user-defined methods. Other features include no-code simulations for diffusion tasks, interactive visualizations, data aggregation, and chart drawing facilities. Designed in Python, Crowd also supports generative agents and connects easily with Python’s libraries for data analysis and machine learning. Finally, we include three case studies to illustrate the use of the framework, including generative agents in epidemics, influence maximization, and networked trust games.
\end{abstract}


\section{Introduction}

\IEEEPARstart{A}{gent-based} modeling and simulation (ABMS) serves as a valuable tool employed for various research fields, such as analysis of adoption of electronic vehicles \cite{evAdoption1}, mobile applications \cite{branchlessBanking}, marketing strategies \cite{marketing1}, disease transmission between animals \cite{insectStudy}, virus spread and intervention strategies for the contagious diseases \cite{covidPBlock, campusStudy}, dissemination of information and trends \cite{socialBots}, trust management \cite{decisionMaking}, transportation planning \cite{lastDelivery}, and emergency evacuations \cite{emergencyEvac}.\par
ABMS employs a bottom-up approach to model real-world systems \cite{abmsurvey2016}, allowing researchers to observe how the behavior of individual entities (agents) influences the system through interactions with other agents and the environment \cite{abmsurvey2022}. By inspecting the agents' actions at the micro level, researchers can explain or predict the emergence of patterns at the system level, such as the diffusion of trends. 

Agents of a simulation include a set of properties and actions defined by the modeler. Depending on the state of the agent and the environment, agents perform these specified actions in each time step of the simulation. In addition, environments can also have properties and, in some simulation tools, their own actions (e.g., patches in NetLogo \cite{netlogo}). Changes in the environment's properties are directly reflected in the selection of agent actions. \par

ABMS has become an essential methodology in computational social sciences, particularly in studying social networks. In agent-based social network simulations, which we simply refer to as social network simulations in the following paragraphs, the environment is represented as a network. Examining the interaction between the agents in this structured environment provides insight into issues such as information diffusion, trust dynamics, and disease spread, which can be leveraged to prevent negative outcomes or achieve positive ones. In a network simulation, an agent typically represents a person, but may also represent an animal, an organization, or any other individual entity. A link between two nodes represents the existence of a relationship between two agents, and there may exist different kinds of relationships (e.g. family, coworker, or possibility of coordination between two organizations). Nodes interact with their neighbors, and the actions and state of these neighbors play a crucial role. For instance,  while modeling the spread of information in simulations, the information is carried to an agent from its neighbors.

Various existing ABMS tools also facilitate the simulation of social networks, with NetLogo \cite{netlogo}, MASON\cite{mason}, Repast\cite{repastSimphony}, and Mesa \cite{mesa} being the most widely used examples. While each framework meets the basic requirements to conduct a network simulation, they differ by the extensiveness of the features in each requirement. The general-purpose ABMS tools are not designed with a focus on social networks. Hence, the extent of features and the effort required to model a complex networked system increases.

In this paper, we introduce \textit{Crowd}\footnote{ \href{https://github.com/bilkent-sna/crowd}{https://github.com/bilkent-sna/crowd}}$^,$\footnote{ \href{https://github.com/bilkent-sna/crowd-ui}{https://github.com/bilkent-sna/crowd-ui}}, a social network simulation framework that aims to simplify and fasten the process of developing agent-based models and simulations on networks. Key features of Crowd include interactive network visualizations, a configuration file approach for setting up simulations, which also allows no-code simulations for diffusion tasks and facilities for merging data and drawing charts for further analysis. While prioritizing ease of use and promoting less code writing, Crowd also allows the incorporation of custom models for more diverse use cases, due to its flexible and extensible architecture. Users of Crowd can take advantage of the Python environment by running their codes on Jupyter notebooks and accessing extensive Python libraries for tasks such as data analysis and machine learning. Although Crowd provides additional features for some use cases, such as the integration of Network Diffusion Library (NDLib)'s \cite{ndlib} compartment structure into configuration settings, it is designed to be a general-purpose tool that can help researchers across various fields. 

Our contributions can be summarized as follows:
\begin{itemize}
    \item We present Crowd, an open-source framework that simplifies the setup and execution of agent-based network simulations through a configuration-based approach.
    \item Crowd supports customization with its extensible structure and allows the execution of user-defined methods written in Python providing access to extensive data analysis and machine learning libraries.
    \item Crowd provides a user-friendly interface that facilitates simulation setup, interactive network visualization, chart generation, and data aggregation, enabling users to model and analyze various research problems.
    \item We demonstrate Crowd's capabilities and ease of use through three case studies: the integration of generative agents into virus transmission modeling, influence maximization, and a networked trust game.
\end{itemize}

In the following sections, we first provide brief information about the previously developed tools for social network simulation and how Crowd compares to them (Section II). Then, we explain the architecture of Crowd (Section III), followed by the simulation development approach, detailing the design, execution, and analysis steps for both the tool and library options (Section IV). We then introduce three case studies to illustrate the use of Crowd, which include the use of generative models for agents' decisions, influence maximization, and a trust game involving investors, trusted investees, and untrusted investees (Section V). We further discuss how to represent common social network parameters with Crowd's features and provide a detailed comparison with Crowd with a general-purpose framework to emphasize its contributions (Section VI). Finally, we conclude with a summary of the framework's key aspects and a discussion of future work (Section VII).

\section{Related Work}
While the general-purpose ABMS tools can be used for a wide range of simulations, the time and effort required to implement a system that can accurately model the real world for complex tasks significantly increases. As a consequence, many researchers develop domain-specific tools tailored to the common fields and topics where ABMS is applied. In this section, we first introduce the general-purpose ABMS tools, then move on to social network simulators, and explain where Crowd is positioned in this mapping. 

To better illustrate this, we present a comparison in Table \ref{tab:comparison}. This comparison is inspired by the classifications made by Antelmi et al. \cite{abmsurvey2022} and Abar et al. \cite{abmsurvey2017}. We evaluate the simulators based on several criteria: programming language for model development, scope of ABMS, available simulation environments, type of graphical user interface (GUI), functionalities for visualization, data collection, batch running, and model exploration. Additionally, we list the effort required for model development and the scalability of each tool, as an extension of Abar et al. \cite{abmsurvey2017}'s classification. 

\begin{table*}
    \centering
    \caption{Comparison of Other Simulation Tools with Crowd}
    \label{tab:comparison}
    \begin{tabular}{|c|c|c|c|c|c|c|c|c|c|c|c|} \hline
        \textbf{\thead{Category}} &
        \textbf{\thead{Tool/\\Property}} & 
        \textbf{\thead{PL for \\Model\\Dev.}} & 
        \textbf{\thead{ABMS\\Scope}} & 
        \textbf{\thead{Simulation\\Environment}} & 
        \textbf{\thead{GUI}} & 
        \textbf{\thead{Vis.}} & 
        \textbf{\thead{Data\\Coll.}} & 
        \textbf{\thead{Batch\\Run}} & 
        \textbf{\thead{Model\\Exp.}} & 
        \textbf{\thead{Model\\Dev.\\Effort}} & 
        \textbf{\thead{Model's\\Scalability\\Level}} \\ 
        \hline 

        \multirow{5}{*}{\makecell[c]{\vspace{1.0cm}\\General-\\Purpose}} &
        \textbf{NetLogo} & 
        \makecell{NetLogo,\\Python*,\\R*} & 
        \makecell{General\\purpose} & 
        \makecell{Grid, network,\\continuous, GIS*} & 
        \makecell{\ding{51}\(^{1,3}\)} & 
        2D, 3D* & 
        \ding{51} & 
        \ding{51}* & 
        \ding{51}* & 
        Simple\(^b\) &
        \makecell{Medium-\\large\(^b\)} \\ \cline{2-12}

        & \textbf{Mason} & 
        Java & 
        \makecell{General\\purpose} & 
        \makecell{Grid, network,\\continuous, GIS*} & 
        \ding{51}\(^{1}\) & 
        2D, 3D & 
        \ding{51} & 
        \ding{51}\(^a\) & 
        \ding{51}* & 
        Complex\(^b\) &
        \makecell{Medium-\\large\(^b\)} \\ \cline{2-12}

        & \makecell{\textbf{Repast}\\\textbf{Simphony}} & 
        \makecell{Java,\\Groovy,\\ReLogo} & 
        \makecell{General\\purpose} & 
        \makecell{Grid, network,\\continuous, GIS} & 
        \ding{51}\(^{1}\) & 
        2D, 3D & 
        \ding{51} & 
        \ding{51} & 
        \ding{51} & 
        Complex\(^b\) &
        Large\(^b\)\\ \cline{2-12}

        & \textbf{Agents.jl} & 
        Julia & 
        \makecell{General\\purpose} & 
        \makecell{Grid, network,\\continuous, GIS} & 
        \ding{55} & 
        2D, 3D & 
        \ding{51} & 
        \ding{51}\(^a\) & 
        \ding{51} & 
        Moderate &
        \makecell{Medium-\\large} \\ \cline{2-12}

        & \textbf{Mesa} & 
        Python & 
        \makecell{General\\purpose} & 
        \makecell{Grid, network,\\continuous, GIS*} & 
        \ding{51}\(^{2}\) & 
        2D, 3D* & 
        \ding{51} & 
        \ding{51} & 
        \ding{51} & 
        Moderate &
        \makecell{Small-\\medium\(^b\)} \\ \hline

       \multirow{4}{*}{\makecell[c]{\vspace{0.55cm}\\Network-\\Focused}} & 
        \textbf{Covasim} & 
        Python & 
        Covid-19 & 
        Network & 
        \ding{51}\(^{3}\) & 
        2D & 
        \ding{51} & 
        \ding{51} & 
        \ding{51} & 
        Simple &
        Large \\ \cline{2-12}

        & \textbf{HashKat} & 
        \makecell{YAML,\\C++} & 
        \makecell{Twitter-like\\networks} & 
        Network & 
        \ding{55} & 
        \makecell{DAT \&\\GEXF\\files} & 
        \ding{51} & 
        \ding{51} & 
        \ding{55} & 
        Simple &
        Large \\ \cline{2-12}

        & \textbf{Soil} & 
        Python & 
        \makecell{Social\\networks} & 
        Network & 
        \ding{51}\(^{2}\) & 
        2D & 
        \ding{51} & 
        \ding{51} & 
        \ding{51} & 
        Moderate &
        \makecell{Small-\\medium} \\ \cline{2-12}

        & \textbf{Crowd} & 
        \makecell{Python,\\YAML}& 
        \makecell{Social\\networks} & 
        Network & 
        \ding{51}\(^{4}\) & 
        2D & 
        \ding{51} & 
        \ding{51} & 
        \ding{51} & 
        \makecell{Simple} &
        \makecell{Small-\\medium} \\ \hline
    \end{tabular}
     \footnotesize{Features marked with an asterisk (*) are available through extensions. \({a}\) and \({b}\) refer to data from surveys \cite{abmsurvey2022} and \cite{abmsurvey2017}, respectively.\\ \({1,2,3,4}\) denote the GUI type: (1) desktop application, (2) local web application, (3) hosted web application, and (4) desktop application built with web frameworks.}

\end{table*}

\subsection{General purpose ABMS tools and frameworks}

\subsubsection{Logo-style simulators}

NetLogo\cite{netlogo} introduces its own programming language, based on Logo, which aims to simplify the modeling process for users with a non-programmer background. It is used to implement the user's own models or to modify a sample model chosen from NetLogo's extensive library, which is directly reachable from the desktop app. With the recent improvement of the large language models (LLMs), the researchers of NetLogo have introduced another study, ChatLogo\cite{chatlogo}, which aims to help new programmers with both starting up their model and fixing the bugs. ReLogo plugin of Repast Simphony\cite{repastSimphony} allows a quick and simple start to modeling, and unlike other Logo tools, it can be mixed with Java and Groovy, providing access to Java libraries. 

\subsubsection{Python-based simulators emphasizing modularity and ease of use}

While Logo-style simulators allow a lower entry barrier to ABMS, this approach requires the users to learn numerous terms for modeling structure. It is not as practical as just using a general-purpose language, such as Python, which is widely used in many disciplines and includes various data analysis libraries in its ecosystem.

Mesa \cite{mesa}, the first Python ABMS framework, provides a modular architecture to support a variety of scenarios. As a consequence, it is considerably easier to define new spaces compared to previous alternatives. Visualization in Mesa is conducted with a web-based UI and is created by users by importing a variety of visualization elements provided for different types of environments in the library.

Another Python framework, AgentPy\cite{agentPy}, emphasizes its ability to provide interactive computing within a single programming environment, Jupyter notebooks. The syntax utilized in AgentPy further simplifies the modeling and parameter exploration, requiring fewer lines of code compared to Mesa. 

NetLogo also allows the integration of Python code in its source code through its Python extension. This enables NetLogo to take advantage of data analysis and machine learning libraries specific to Python. Python code is passed as a string to the call function, and the  results returned from Python can be utilized for visualization of the model and generation of charts. While this approach eliminates a limitation of NetLogo, regarding access to popular Python libraries, it is impractical compared to using a Python-based simulator like Mesa or AgentPy. This development environment does not provide any extensive support for Python in terms of code completion, debugging, and other features included in IDEs like VSCode, or interactive code execution that comes with Jupyter notebooks. 

\subsubsection{Larger scale simulators}

As the number of agents in a simulation increases to hundreds of thousands or millions, execution times significantly increase. To address this issue, various tools provide performance optimizations with multithreading, distributed computing, or tensor-based approaches. 

Besides the execution speed, MASON\cite{mason} focuses on excluding any platform dependencies. It adopts a three-layer design approach—utility, model, and visualization—and facilitates model checkpointing to stop and continue execution, which allows switching machines. Within the Repast family of ABMS tools, Repast Simphony (Java)\cite{repastSimphony} emphasizes a plug-in architecture for modularity, and provides features such as logging simulation information at runtime, data aggregation, customizable shapes for visualization, and distributed execution. Repast HPC (C++)\cite{repastHPC} allows parallel distributed runs of simulations designed to be executed on high-performance computers. Repast4Py (Python)\cite{repast4Py} aims to lower the entry barrier for modelers to try distributed simulators by simplifying the process. DeepABM\cite{deepABM} stores each agent and their states in tensors and models the interactions between them using graph neural networks (GNNs).

Recent additions to performance-oriented simulators include Agents.jl\cite{agentsjl} and krABMaga\cite{krABMaga}. Agents.jl leverages Julia's performance and ease of defining complex mathematical expressions. It adds new features to the ABMS space, such as N-dimensional space support and agent sampling based on certain properties. krABMaga takes advantage of Rust's efficiency and reliability. Besides the core simulation functionalities, it facilitates model explorations in parallel, distributed, and cloud architectures. 

\subsection{Social network simulation tools and frameworks}

Social network platforms such as Facebook, X (Twitter), and Reddit enable the quick dissemination of information without the location barrier. Consequently, these platforms become a place to spread misinformation or trends for business purposes, which has been of interest to many researchers. The inspection of the effects of certain people on the general public aligns closely with the methodology of ABMS, where the influence of individual entities on the system can be analyzed. 

Hashkat\cite{hashkat} allows a detailed analysis of Twitter-like large-scale networks with its built-in functionalities, including the generation of networks with agent properties. HashKat requires users to create a configuration file to define the simulation properties, such as time limitation, follow models, and output settings (e.g., visualization, tweet analysis).  Despite providing extensive support for realistic social network simulations, it requires more effort to extend the simulator's capabilities compared to modular simulators. The configuration aspect is similar to our approach; however, our tool is designed to be more general-purpose and extensible.

The Open Source Social Network Simulator \cite{spatialSNSim} focuses on spatiotemporal aspects of information diffusion. It provides simple network generation, Independent Cascade and Linear Threshold diffusion models, commonly used social network analysis (SNA) metrics, and algorithms for community detection and selecting seed nodes within its GUI. With its text analysis tool, users can process files to access keyword usage information and plot graphs of word frequency or shifts over time. Contour maps can be generated to compare simulations with real-world data. While Crowd also simplifies the model implementation by allowing no-code diffusion simulations, it still requires the user to set the rules. Although this approach requires more effort, it gives the user the flexibility to define other diffusion models more easily.

Soil\cite{soil} is an extensible Python-based simulator, designed to work with Mesa in the later versions to allow its users to take advantage of both libraries. The tool provides network, agent, and event classes, and a web-based visualization of the network, settings, and charts. Soil utilizes Python to describe the simulation, SQLite database to store the results, and YAML files for simulation and visualization configurations.

With its Python environment, configuration file, network generation, and basic visualization approaches, Soil is the most similar study to Crowd. However, these frameworks differ in the simulation definition and execution methodologies, visualization capabilities, and ease of use. Crowd integrates compartment structure into its configuration style, a default data collection system for graphs and user-implemented methods, and a Tauri-based cross-platform GUI for project navigation, simulation definition, and execution, all through the user interface. All of these features lower the entry barrier for agent-based network simulations while requiring less time for coding at each step.

Social networks are also utilized in epidemiological studies, such as the analysis of the spread of diseases like COVID-19, where modeling the dissemination of the virus, understanding the impact of different intervention strategies, and estimating resource requirements are typically the main objectives. To significantly reduce the time required to model the effects of these factors in a simulation environment, tools such as Covasim \cite{covasim} and Flu and Coronavirus Simulator (FACS) \cite{facs} have been developed. 

\section{Architecture}

\begin{figure}[!t]
    \centering
    \includegraphics[width=0.87\linewidth]{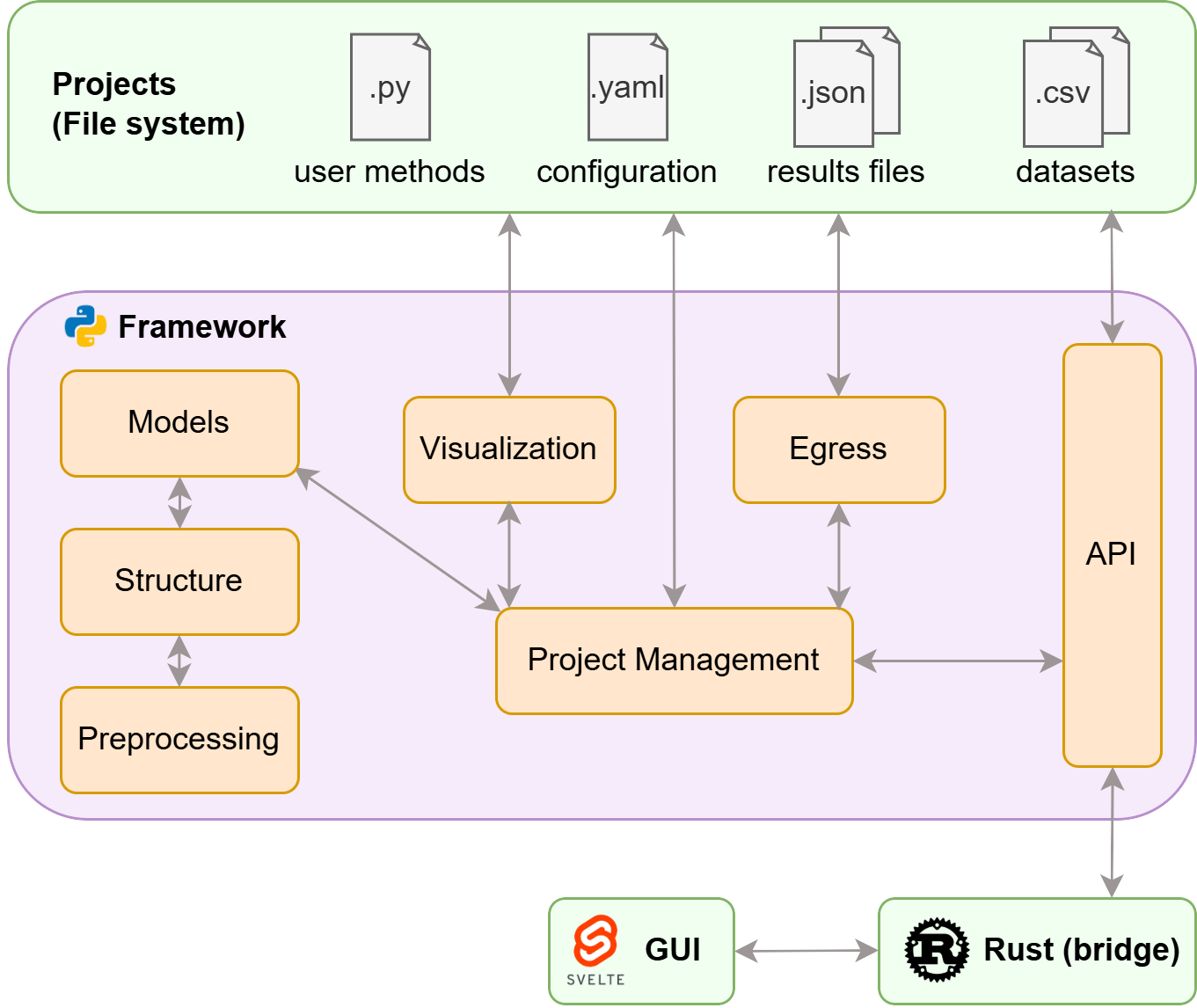}
    \caption{Architecture of the Crowd framework.}
    \label{fig:architecture}
\end{figure}

We developed Crowd as a Python library that provides facilities for social network simulations, with extensible modules designed to include the fundamental functionalities for the task, such as network models, data savers, visualizers, and the execution of simulation. We also developed a desktop application that provides an extra layer to the library to increase the ease of use and take advantage of the JavaScript visualization libraries. We demonstrate the general architecture of the Crowd framework, illustrating how the library connects to the desktop app and the file system, as shown in Fig. \ref{fig:architecture}.

With this app, users can run simulations by selecting the settings and parameters, view the previous simulations of a project, and navigate between projects and simulations in a user-friendly fashion. For more complex simulations and data collection purposes, users can write Python methods within an editor and select when these methods will be executed in the simulation life cycle. After a simulation is executed, users can explore the results with graph visualization and node and edge information sections, aggregate data from multiple simulations, and draw charts using simple selectors.  

Instead of requiring the researchers to write code to set up a simulation, Crowd utilizes YAML files to describe network settings, node types, node and edge parameters, and in the case of diffusion simulations, compartments and the rules. This allows no-code simulations for simple models, while notably reducing the lines of code needed for others. While using the library version of Crowd, the features that are not provided by this setup can be added by the modeler and executed before calling the library method that runs the simulation. An example of this use can be initializing an LLM before passing it to be used for inference in each step of the simulation. 

 To increase ease of installation and make the tool more approachable for non-technical users, we utilize Tauri to provide a cross-platform application. The Rust layer acts as a bridge, relaying function calls from the GUI (Svelte) to the backend (Python). The functions that can be called with this structure are placed in another module called Crowd API, and these functions facilitate the execution of core simulation tasks and management of file operations, such as uploading datasets and accessing project data and results stored as JSON and YAML files in the file system. 

\subsection{Network architecture}
\label{subsec:network_architecture}

In the other ABMS frameworks, such as Mesa and MASON, two classes are expected to be implemented by the user, Model and Agent. Crowd adopts a different approach. In the current version, the only environment that the simulation can reside on is a network. This network environment consists of nodes and edges, where nodes are the agents and edges define the relations between the agents. The information related to agents is stored as node parameters in the network, and agent-level methods can be defined separately to pass as a list parameter to the library method that runs the simulation. 

We provide two base classes, \texttt{Network} and \texttt{CustomSimNetwork}, as well as two extensions of them: \texttt{DiffusionNetwork} and \texttt{EdgeSimNetwork}. To represent networks in our framework, we utilize NetworkX \cite{networkX}, which is a library that provides data structures to represent graphs, along with algorithms for social network analysis and the generation of networks. For the network generation task, we also use the igraph-python library\cite{igraph-python}. 

The base \texttt{Network} class of Crowd is designed to provide a general framework for all possible user-defined scenarios. For the design of this network type, we adopt a \textit{network-as-code} approach where the network, agents, their interactions, details of the analysis, and visualization are defined using configuration files and code. It is useful for researchers who want to implement all low-level details of the simulation. We extend this network type for a use case of edge-based simulation where we add links to the network in every iteration of the simulation, depending on the update method provided by the user for the selection. This class, which we call \texttt{EdgeSimNetwork}, illustrates the extensible nature of Crowd and how data savers and visualizers can be utilized in custom networks created by the modelers. 

The third network type called \texttt{CustomSimNetwork} provides notably more functionalities compared to previous types. \texttt{CustomSimNetwork} handles the addition of network, node, and edge parameters, and holds and runs user-defined methods in the specified times. \textit{Before-iteration} methods allow setting or resetting some parameters before the main logic of the iteration. An example of this can be a person (agent) deciding to go out or stay home depending on the number of COVID cases in their town and updating their location parameter accordingly (case study 1). Between\textit{ before-iteration} and \textit{after-iteration} methods, functions labeled as \textit{every-iteration-agent} are executed, where we call the given methods for each agent in random order. \textit{After-iteration} methods can be utilized for data collection purposes, such as finding the total number of agents deciding to stay home each day. Moreover, \textit{after-simulation} methods can provide insight into the final situation of the case or finalize any remaining logic. 

\texttt{DiffusionNetwork}, an extension of the \texttt{CustomSimNetwork} class, facilitates simplified diffusion simulations where the simulation logic is defined using the compartment structure from NDLib. This allows NDLib to provide all functionalities related to the diffusion, while the modeler needs to only write methods for data collection purposes.

 \subsection{Data collection, batch runner and model exploration}

Collection of the simulation data is crucial for visualization and analysis at different timestamps. The extent of automated data collection in Crowd varies depending on the selected network type, while writing the state of the NetworkX graph is a common feature for all.  The simulations based on the \texttt{CustomSimNetwork} store the node type counts and changes in them, as well as the results of the user-implemented methods. If a user-implemented method returns a value, it will be saved in a JSON file with the name of the function. This file will then be available to use for further data analysis, which can be conducted by drawing charts on the GUI or loading the data into a dictionary or a pandas DataFrame with Python.  

The user-defined methods can take advantage of the Python ecosystem by importing and utilizing any other library the user has installed on their machine, both in the tool and library versions. This includes but is not limited to NumPy, pandas, scikit-graph, and machine learning frameworks and libraries, such as TensorFlow, PyTorch, Scikit-learn, or Keras. This highlights another benefit of using Python for modeling, compared to Java-based simulators. Furthermore, modelers can utilize the extensive functionalities that NetworkX provides for social network analysis, including the network and node level metrics and community detection algorithms. 
 
As simulations may include some random processes, many ABMS tools, including Crowd, provide batch runners. With this capability, the simulations are repeated with the same parameters, and the results are aggregated, commonly by taking the average of the collected data. Model exploration, or parameter sweep, allows modelers to experiment with different parameter values. In Crowd, the list of values to explore can be given in the related section of the configuration file. Crowd generates combinations of simulation settings where only one value is tested at a time, and executes all these configurations. The resulting data can then be used to identify optimal parameter values or inspect the impact of an independent variable on the dependent variables. The data of these simulations can also be merged, with the restriction of having the same file name and content format. 

\section{Simulation Development Approach}

\begin{figure}[!h]
    \centering
    \includegraphics[width=0.95\linewidth]{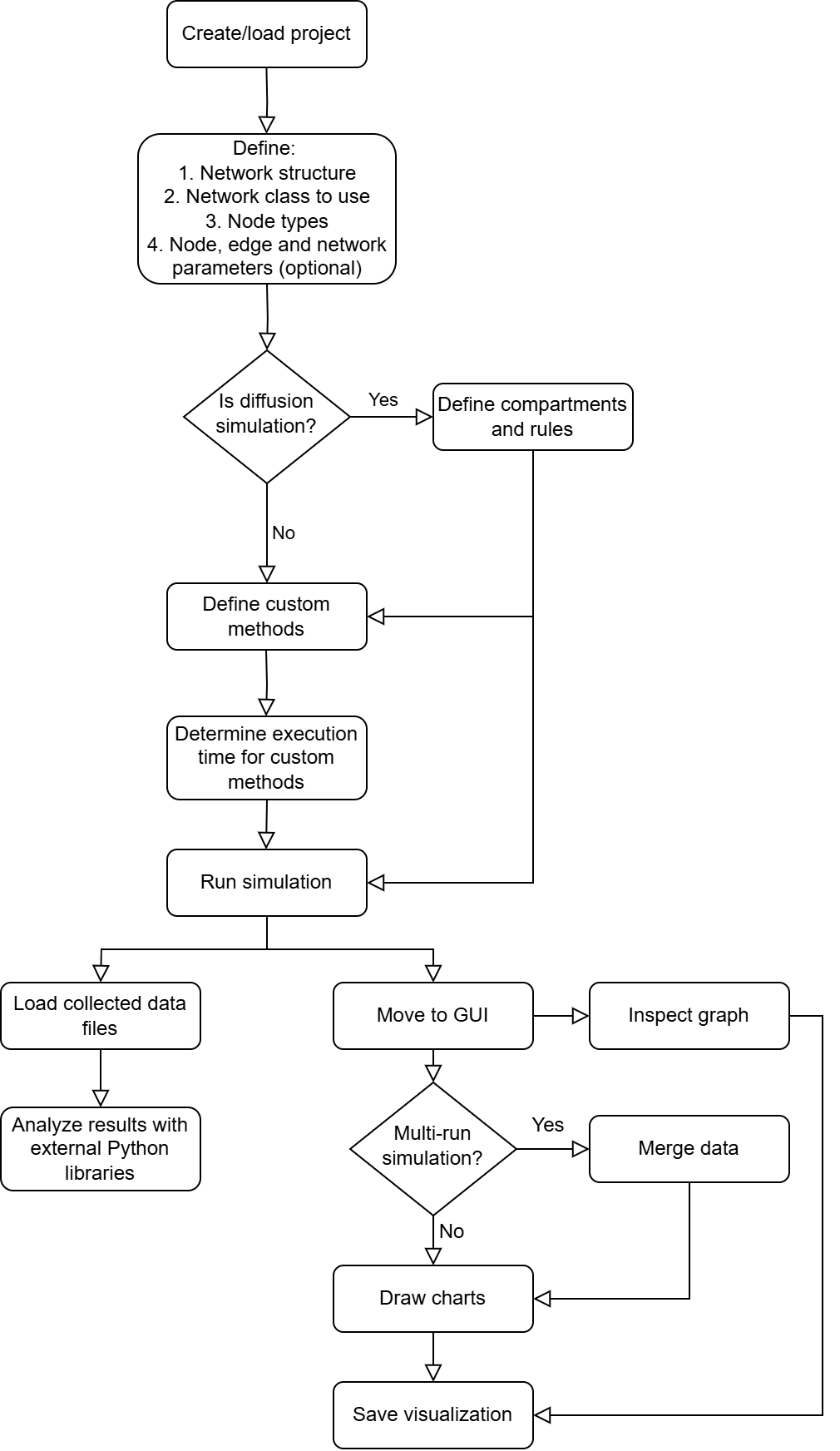}
    \caption{Simulation development and analysis steps of the Crowd framework.}
    \label{fig:flowchart}
\end{figure}

In this section, we describe the steps to design, run, and analyze a simulation utilizing the library and tool versions of Crowd in detail, while also providing a summary in Fig. \ref{fig:flowchart}. We chose the widely used Susceptible-Infected-Recovered (SIR) model for a basic diffusion simulation that can demonstrate the distinct features of Crowd. The mathematical formulation and details of the SIR model are provided in \cite{SIR_model}.

\subsubsection{Step 1: Create or load project}

In Crowd, simulation settings, datasets and results are encapsulated in a Project structure. To begin a simulation process, a project should be either created or loaded using the library's functions or the GUI.

\subsubsection{Step 2: Modify configuration}

The configuration file approach of Crowd allows all simulation settings to be defined in a simple way. The structure component describes how the network will be initialized, which can be from a file or randomly generated with various available methods. In Listing \ref{lst:structure-yaml}, we select our network to be a random regular network with 100 nodes, having a constant degree of 4. 

\begin{lstlisting}[language=YAML, caption={Defining network structure in configuration file.}, label={lst:structure-yaml}, captionpos=b]
name: SIR-example
structure:
  random:
    count: 100
    degree: 4
    type: random-regular
\end{lstlisting} 

Alternatively, modelers can also use Crowd's GUI for simulation configuration with buttons and dropdowns, and the selections are automatically converted into YAML.

The \textit{definitions} component is where the properties for network initializations reside. In diffusion simulations, compartments and rules are also included in this section. In Listing \ref{lst:nodetypes-yaml}, we first specify the network type to use, then list the node types and their initialization methods. As explained in the previous section, the agent and model parameters are stored in the network object. In this file, these parameters can be defined to get initialized randomly with the given range or options. Compartments are rule bits that specify the conditions of a node state change. NDLib provides several types of compartments that can be utilized with different combinations for various simulations. In cases where these compartments are not enough to describe a problem, modelers can switch to custom simulation networks and include their own methods for the task. 

\begin{lstlisting}[language=YAML, caption={Definitions for SIR example with Diffusion Network.}, label={lst:nodetypes-yaml}]
definitions:
  pd-model:
    name: diffusion    
    nodetypes:
      Susceptible:
        random-with-weight:
          initial-weight: 0.9
      Infected:
        random-with-weight:
          initial-weight: 0.1
      Recovered:
        random-with-weight:
          initial-weight: 0
    node-parameters:
      numerical:
        age:
          - 0
          - 100    
    compartments:
      c1:
        type: node-stochastic
        ratio: 0.1
        triggering_status: Infected
      c2:
        type: count-down
        name: healing
        iteration-count: 4
    rules:
      r1:
        - Susceptible
        - Infected
        - c1
      r2:
        - Infected
        - Recovered
        - c2
\end{lstlisting}

\begin{figure}
    \centering
    \subfloat[]{
        \includegraphics[width=0.5\linewidth]{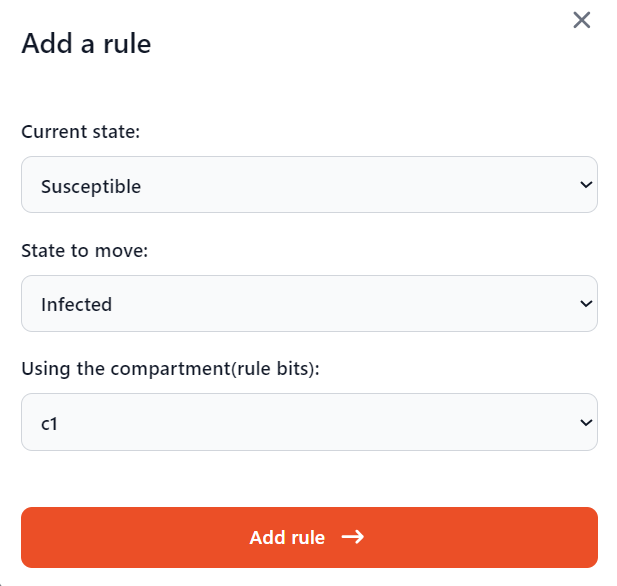}
        \label{fig:add-rule}
    }
    \hfill
    \subfloat[]{
        \includegraphics[width=0.41\linewidth]{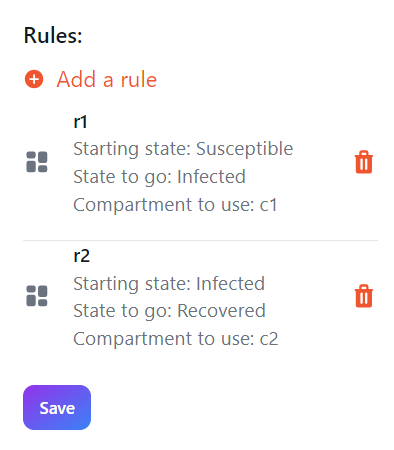}
        \label{fig:view-rules}
    }
    \caption{Rule addition in the Crowd GUI for the SIR example. (a) Adding a rule. (b) Viewing the rules.}
    \label{fig:rules-ui}
\end{figure}

The first rule that uses compartment \textit{c1} included in Listing \ref{lst:nodetypes-yaml} runs as follows: If a node is susceptible, execute \textit{c1}. If this node has any  Infected neighbors (triggering status),  with a probability of 0.1 (ratio), the node switches from Susceptible to Infected state. Rule 2 describes the Infected to Recovered sequence with a countdown compartment to define that after 4 iterations, the node will switch states. Additionally, the process of constructing these rules in the GUI is illustrated in Fig. \ref{fig:rules-ui}.

\subsubsection{Step 3: Define custom methods and run simulation}
\label{subsubsec:sim_dev_steps_def_methods}

In cases where additional methods are needed either to facilitate the logic specific to that simulation or data collection, users can describe their own methods in the Method Lab section of the GUI. Any Python package that is installed on a user's computer can be imported and used in these functions. We provide a code editor and selectors to specify the stages of the simulation during which each function will be executed.

The library equivalent of this process is given in Listing \ref{lst:run-simulation}.  Every function that will run during or after the simulation has access to the network object, and it should be included as a parameter while writing custom methods. Additional information about important points of defining a custom method is explained in Crowd's documentation\footnote{\href{https://crowd.readthedocs.io}{https://crowd.readthedocs.io}}. 

\begin{lstlisting}[language=Python, caption={Define custom methods and run simulation.}, label={lst:run-simulation}]
# Returns the % of infected nodes in every snapshot
def get_percentage_infected(network):
    return (network.node_count[1] /
           network.G.number_of_nodes()) * 100

# Run the simulation
my_project.lib_run_simulation(epochs=50,
                              snapshot_period=5,
                              curr_batch=1,
                              after_iteration_methods=[
                                get_percentage_infected
                              ])

\end{lstlisting}

We want our data collection method to be executed after diffusion for the current iteration is completed; therefore, we pass it as an \textit{after iteration} method. The numbers we return in each iteration are saved by default. We run the simulation for 50 iterations (epochs), save the graph and data collectors every 5 iterations (snapshot period), for one time (curr batch number).  

\subsubsection{Step 4: Analysis}

\begin{figure*}[h!]
    \centering
    \subfloat[]{
        \includegraphics[width=0.22\linewidth]{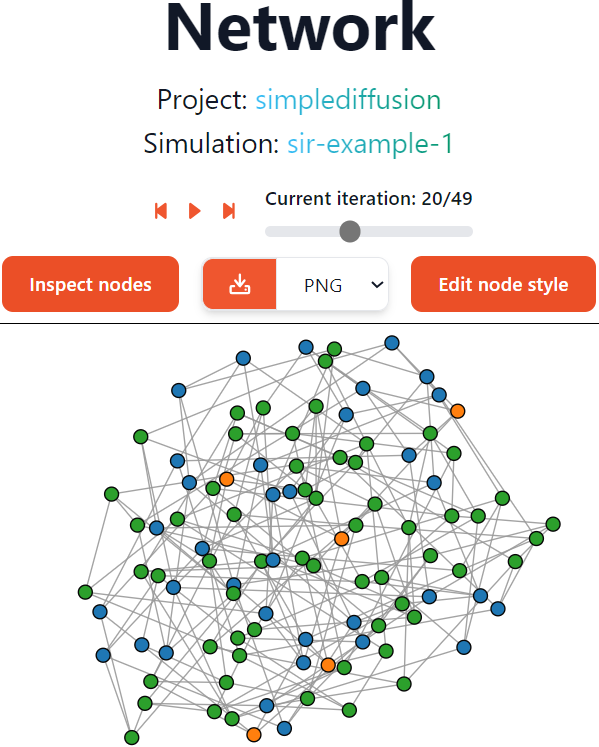}
        \label{fig:network-tab}
    }
    \hspace{0.12\linewidth}
    \subfloat[]{
        \includegraphics[width=0.55\linewidth]{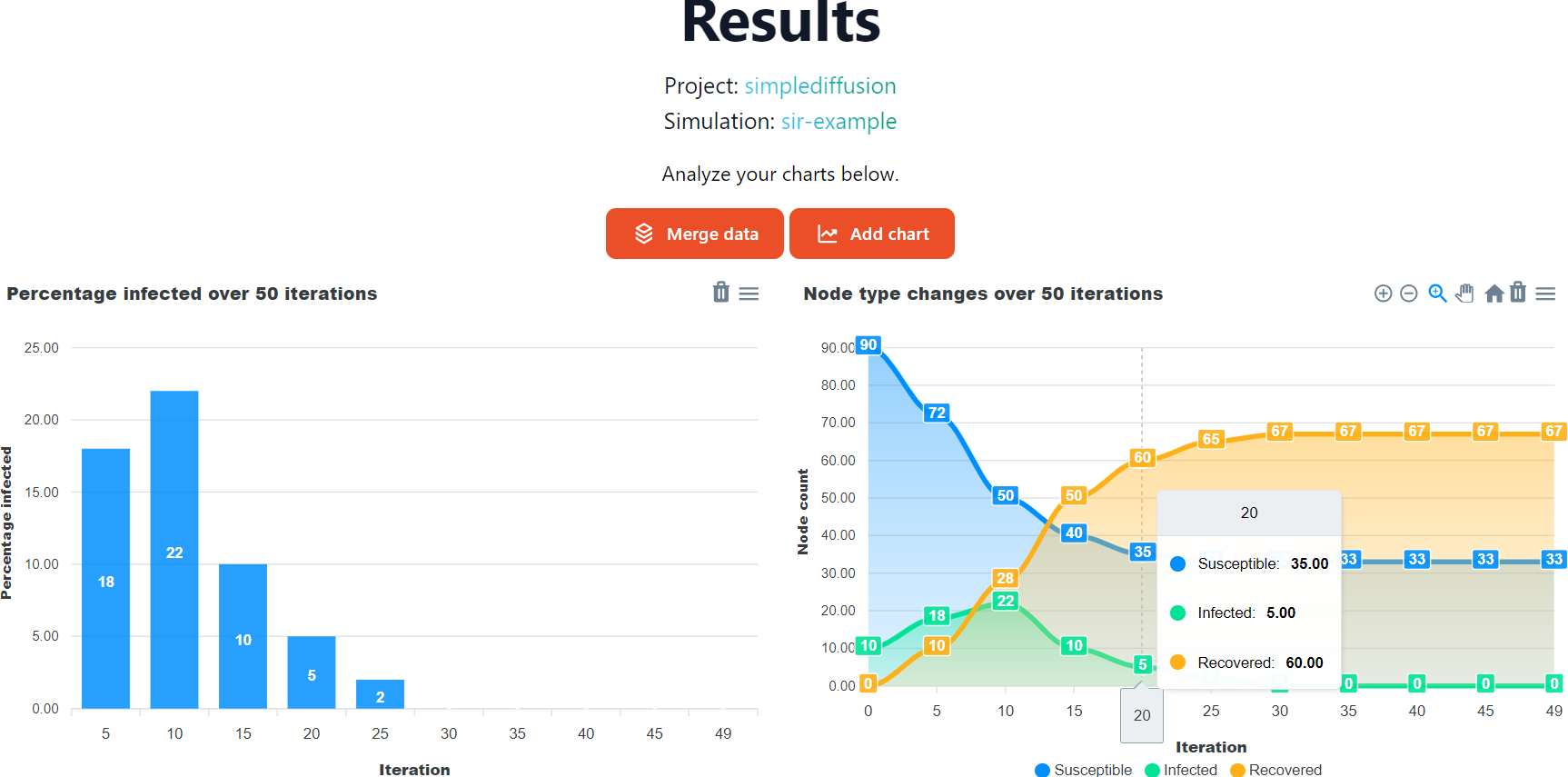}
        \label{fig:results-tab}
    }
    \caption{Network and Results sections of Crowd's GUI. (a) Network tab. (b) Results tab.}
\end{figure*}

Following the completion of the simulation, we move to the Network tab of the GUI where we can observe the changes in the graph, saved every snapshot period in the previous step, as illustrated in Fig. \ref{fig:network-tab}. The network visualization is interactive; it can be moved, zoomed in or out, and node visualization properties can be changed. For more advanced visualizations, the network can be exported in the GEXF format to be loaded to Gephi. 

As mentioned in the previous section, data merge and chart drawing operations can be conducted in the Results tab (Fig. \ref{fig:results-tab}). The figures drawn this way are also interactive. The bar chart uses the data collected with our custom method, while the area chart uses the auto-collected node type counts. Besides the visualization, the researchers can use the JSON files saved in the simulation folder for further data analysis. 

\section{Case Studies}
In this section, we intend to illustrate how Crowd simplifies and streamlines the implementation of existing research across diverse and complex domains by presenting practical examples through various use cases. The first scenario describes the inclusion of generative agents in the simulation of COVID-19 spread, an implementation of the work by William et al. \cite{gabmEpidemic} in a network environment. In our second scenario, we explain how Crowd can be utilized for influence maximization with a real social network from Facebook \cite{facebook_dataset}, commonly used in the field. Finally, we describe the implementation of a trust game scenario between the investors and trustworthy/untrustworthy trustees, changing their strategies according to the payoffs received by their neighbors, based on the paper by Chica et al. \cite{nPlayerTrust}. 

\subsection{Scenario 1: Generative Agents in Social Simulations}

% \begin{figure*}
%     \centering
%     % First chart
%     \begin{minipage}[c]{0.49\linewidth}
%         \centering
%         \includegraphics[width=.8\linewidth]{llm_figures/llm_stay_home.pdf}
%         \subcaption{}
%         \label{fig:stay-home}
%     \end{minipage}
%     \hfill
%     % Second chart
%     \begin{minipage}[c]{0.49\linewidth}
%         \centering
%         \includegraphics[width=.8\linewidth]{llm_figures/llm_4th_day.pdf}
%         \subcaption{}
%         \label{fig:infected-llm}
%     \end{minipage}
%      \caption{Charts generated with Crowd for the generative agents experiment.}
% \end{figure*}

\begin{figure*}
    \centering
    \subfloat[]{
        \includegraphics[width=0.39\linewidth]{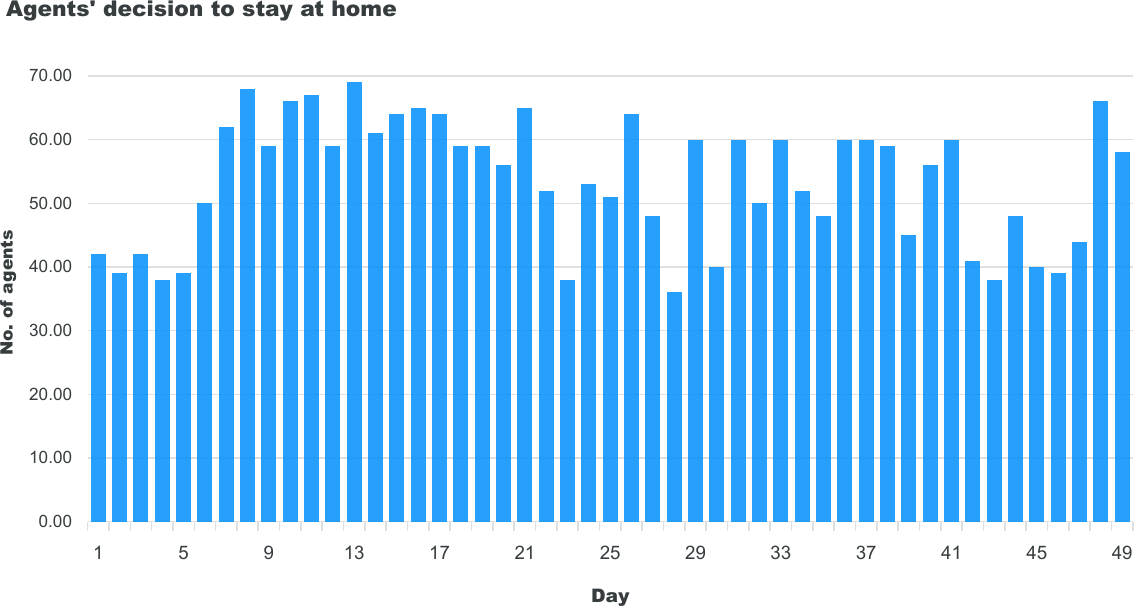}
        \label{fig:stay-home}
    }
    \hspace{0.10\linewidth}
    \subfloat[]{
        \includegraphics[width=0.39\linewidth]{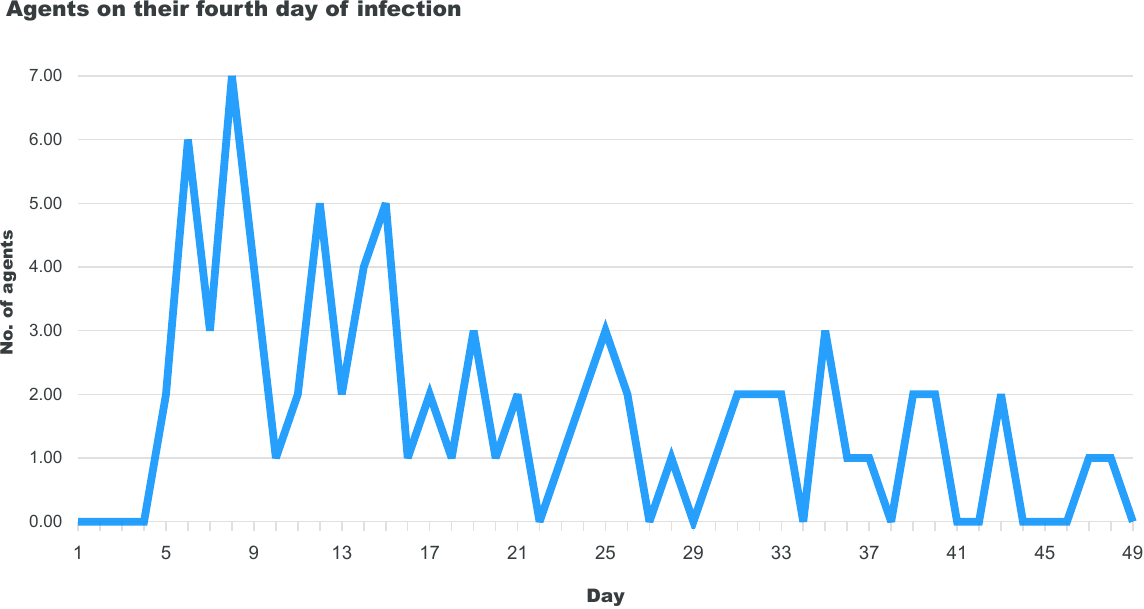}
        \label{fig:infected-llm}
    }
    \caption{Charts generated with Crowd for the generative agents experiment. (a) Agents staying at home. (b) Agents on the fourth day of infection.}
\end{figure*}

In social simulations, agents commonly represent people in real life. However, these agents do not possess the common knowledge humans have, and their decision-making processes are restricted by the information provided by the modeler.
 Recently, researchers have addressed this challenge by integrating large language models (LLMs) into agent-based models, utilizing their capabilities such as world knowledge and common sense reasoning for agents' decisions \cite{llmBasedSurvey}.  While the LLM agents are utilized in social simulation to model virtual towns and communities \cite{agentSims} and to develop simulators \cite{s3}, this approach has also been adopted in the fields of economy, natural sciences, and engineering\cite{llmBasedSurvey}.  A study in the field of epidemiology, GABM-Epidemic \cite{gabmEpidemic}, shows that when given information about the daily cases, agent's characteristics, and infection symptoms, LLM agents behave similarly to real people. In this section, we present an implementation of this study in the Crowd framework to demonstrate its utility as a Python library. It seamlessly integrates with other libraries, such as Hugging Face Transformers for LLM inference tasks, within environments like Jupyter notebooks.

The distinctive point of this study from a standard SIR model is that at the beginning of each day (iteration), each agent decides whether to stay home or go out. If an agent goes out, they become a part of the epidemic network and interact with the other agents. On the other hand, if an agent stays home, they cannot get infected or infect other people. This decision is made by sending a prompt to the LLM, which is constructed according to the agent's age, personality traits, and memories. Memories of an agent consist of their symptoms, information about the virus spread, and their reason to go out, which is defined as work for all agents. The prompt is concluded by asking the model if the agent should stay home. The output provided by the LLM includes a short answer, ``yes" or ``no," and a reasoning of one to two sentences. In a rule-based simulation, agents' personalities do not impact their decision to go out; however, with this approach, a person described as reckless may decide to go out regardless of the high number of daily new cases. 

In the original study, agents reside on a grid environment without a network structure, and the agents to interact with are chosen randomly for each agent, up to five contacts. As Crowd uses a network environment for simulations, all agents are placed on a random regular network, where each node has the same degree. In this setup, each node is connected to five other nodes. In each simulation step, agents interact with their neighbors instead of random agents. Due to these modifications in simulation settings, a direct comparison of simulation results may not be meaningful.

Moreover, William et al. \cite{gabmEpidemic} utilize ChatGPT-3.5 \cite{chatgpt} for the task, while we chose the Mistral 7B – Instruct model \cite{mistralLLM} due to its open-source nature and its superior performance over other models with similar parameter sizes in the common-sense reasoning task, as observed at the time of implementing this case. We conduct the experiment on Google Colab using a T4 GPU. The model is imported from the HuggingFace transformers library with 8-bit quantization applied for faster inference and reduced costs. With each query to LLM taking approximately 10-10.5 seconds, we complete a single experiment of 50 days and 100 agents, consisting of 5000 queries, in 14 hours 17 minutes (857 minutes). 
 
In the configuration file, we first define the node types (susceptible, infected, recovered) and then assign the node parameters (location, personality traits, and age).  

\begin{lstlisting}[language=YAML, caption={Compartments in the generative agents simulation.}, label={lst:llmcompartments} ]
compartments:
      c1:
        type: node-categorical
        attribute: location
        probability: 0.1
        value: grid
      c2:
        type: count-down
        name: healing
        iteration-count: 6
\end{lstlisting}

For diffusion simulations that can be defined using NDLib's compartment structure, the rules can be defined in this file without the need for any code for the basic logic of the diffusion process. The rules of this simulation can be written with only two compartments given in Listing \ref{lst:llmcompartments}. \textit{c1} is a node categorical compartment, which means that the node will move from state 1 to state 2 if the categorical condition is satisfied. Hence, \textit{c1} can be translated as the following:\textit{ If a node's location is grid, with 0.1 probability, Susceptible nodes will transfer to Infected.} When a node is infected, it takes 6 days to recover. We implement this using a count-down compartment (\textit{c2}), which holds a variable called \textit{healing} for the infected agent. When the agent is infected, healing starts with the value of 6, and it decreases by 1 every day. When this number reaches 0, the agent will go to the Recovered state. 

\begin{figure}[!h]
    \centering
    \includegraphics[width=0.8\linewidth]{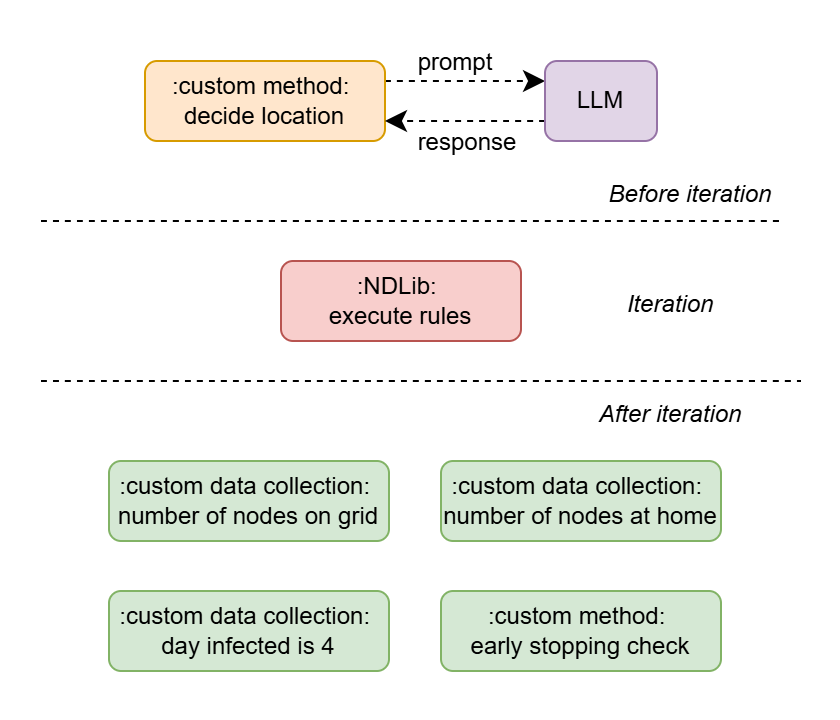}
    \caption{Execution order of a full iteration in the generative agents experiment.}
    \label{fig:llm-iteration}
\end{figure}

Additional methods are implemented to execute before and after each iteration, which are passed to Crowd's run method as list parameters.As shown in Fig. \ref{fig:llm-iteration}, determining the location using prompts sent to an LLM is performed prior to executing the infection logic. Statistical methods are run after the iteration is completed. If these methods return a value, it is automatically saved to a file in every snapshot period, which can be used for analysis in the later steps. 

Fig. \ref{fig:stay-home} is an example bar chart drawn by selecting this automatically saved data in the results page of GUI, which illustrates the number of agents deciding to stay home each day of the simulation. New daily active cases are recorded using a data collection method that counts the number of agents on their fourth day of infection. The percentage of new daily cases relative to the total population is included in the prompt, influencing the LLM agent's decision to stay home or go out. As shown in Figures \ref{fig:stay-home} and \ref{fig:infected-llm}, the rises and falls in both graphs follow a similar pattern.

\begin{figure}[!h]
    \centering
    \subfloat[]{
        \includegraphics[width=0.47\linewidth]{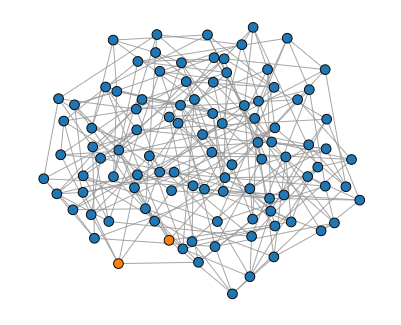}
        \label{fig:llm-initial}
    }
    % \hfill
    % \hspace{0.02\linewidth}
    \subfloat[]{
        \includegraphics[width=0.47\linewidth]{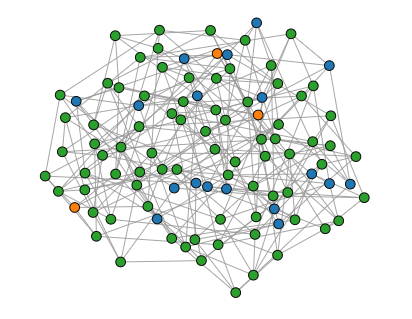}
        \label{fig:llm-final}
    }
    \caption{Network visualization of the generative agents experiment: $n=100$ and $R_0 = 3$ (blue: Susceptible, orange: Infected, green: Recovered). (a) Initial setting. (b) After 50 iterations.}
    \label{fig:llm-network-viz}
\end{figure}

Previously mentioned analysis methods allow us to inspect the simulation from a cumulative point of view. For a node-specific view, we can utilize the interactive network visualization (first and last iterations given in Fig. \ref{fig:llm-network-viz}) or the \textit{inspect node} tab of the GUI, where the values of node parameters and neighboring nodes are listed. 

For simulations employing Generative Agent-Based Models (GABM), DeepMind Concordia \cite{concordia} is an alternative to Crowd. Besides the actors of the scenario being represented by LLMs, the environment in Concordia is also represented and managed by an LLM. This approach greatly differs from rule-based simulations, where the rules defined by the modeler have full control of the simulation. While Concordia supports programmed logic for agent state updates, frameworks like Mesa are more practical when rule-based decision-making is the main methodology. 

Although Crowd and Concordia are both Python-based simulation frameworks, they differ mainly in their focus (social networks vs. GABM) and simulation environment (network-based vs. LLM-managed). Crowd connects agents with links and allows interactions with the other agents (e.g., their neighbors) in a structured environment. Concordia's simulations are conducted within a generative environment, where the agents to interact are determined based on the scenario rather than the underlying network topology. 

Agent information in Concordia is represented as strings stored in an associative memory structure. While Crowd does not provide a built-in agent memory structure, modelers can implement and integrate one as needed. Moreover, Crowd provides more visualization facilities, including interactive network visualization, node and edge inspection, and data merging for drawing charts, whereas Concordia mainly supports chart generation.

Overall, Crowd eliminates the need to write any code for infection logic and visualization tasks for this study, only leaving the task-specific LLM prompting and data collection to the modelers. Furthermore, the code structure of this case study serves as a template for future studies employing generative agents for simulations within the framework.

\subsection{Scenario 2: Influence Maximization}
Influence maximization (IM) in social networks is vastly studied with a wide range of applications in economy, politics, and social sciences. The motivation of this task is to find a set of nodes that can spread information to the maximum number of users in a given network \cite{infMaxSurvey}. The selected (seed) nodes are marked as ``active", and they try to activate the other nodes that were marked as ``inactive" in the beginning. In various studies, researchers combine their selection of seed nodes with a diffusion simulation with Independent Cascade (IC) \cite{ICModel} and Linear Threshold (LT) \cite{LTModel1, LTModel2} models, to show the effectiveness of their novel algorithms. This often requires a comparison with the other available methods, and running the simulation multiple times. Within our framework, we aim to simplify and fasten this process by allowing various initial node type selections and automated data collection. Besides selecting the nodes randomly with count and weight input, modelers can determine the top k nodes with PageRank, degree, in-betweenness, closeness, eigenvector, and Katz centralities. Reading from a file option is especially useful to load the seed nodes determined with the researcher's own algorithm in the previous stage of the study or the other methods to compare that are not provided by Crowd.   

\begin{figure*}[h!]
    \centering
    \subfloat[]{
        \includegraphics[width=0.35\linewidth]{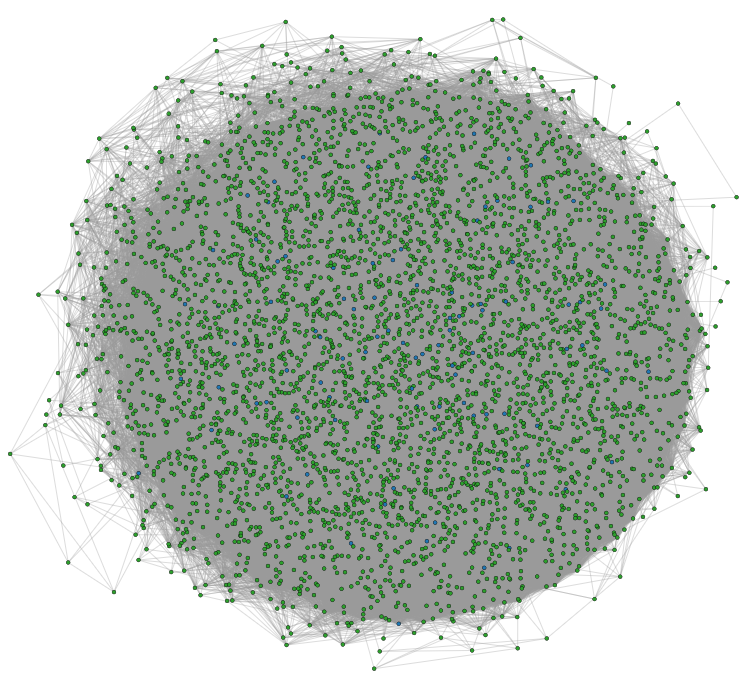}
        \label{fig:page-rank-initial}
    }
    \hspace{0.04\linewidth}
    \subfloat[]{
        \includegraphics[width=0.35\linewidth]{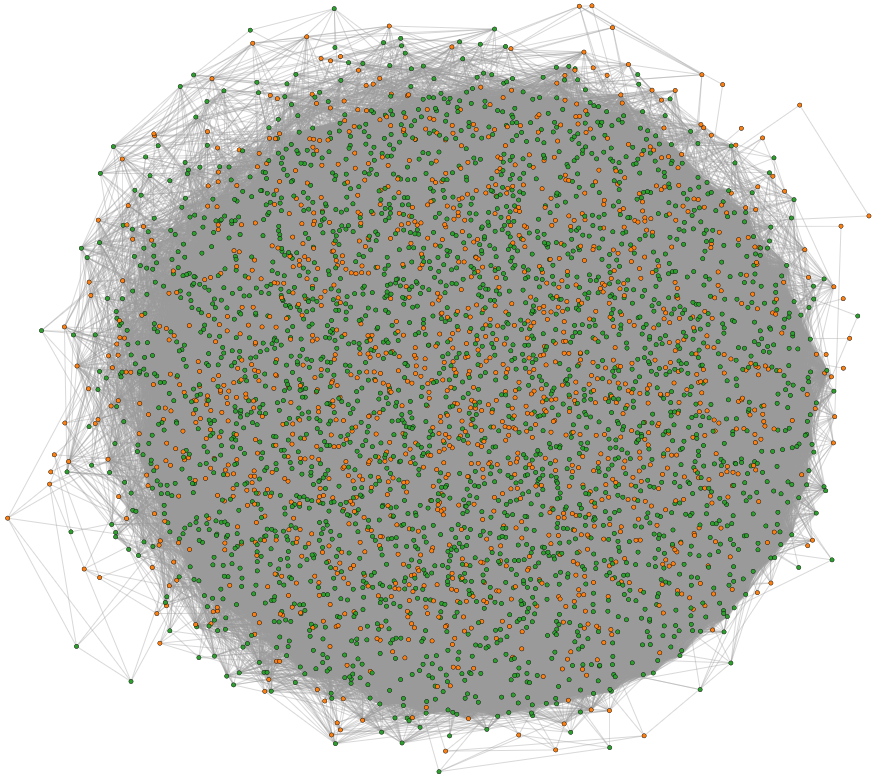}
        \label{fig:page-rank-final}
    }
    \caption{Network visualization of the IM experiment: $n = 4039$, $k = 100$, method = PageRank (blue = Active Spreader, green = Inactive, orange = Active). 
    (a) Initial setting. 
    (b) After 20 iterations. }
    \label{fig:infmax-network}
\end{figure*}

\begin{figure*}[h!]
    \centering
    \subfloat[]{
        \includegraphics[width=0.4\linewidth]{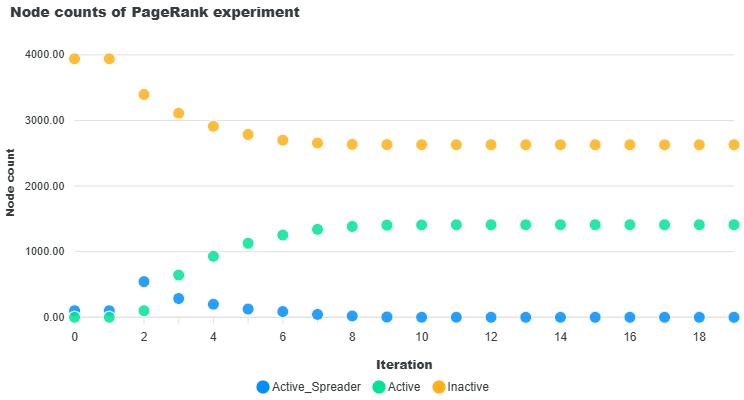}
        \label{fig:page_rank_scatter}
    }
    \hspace{0.04\linewidth}
    \subfloat[]{
        \includegraphics[width=0.4\linewidth]{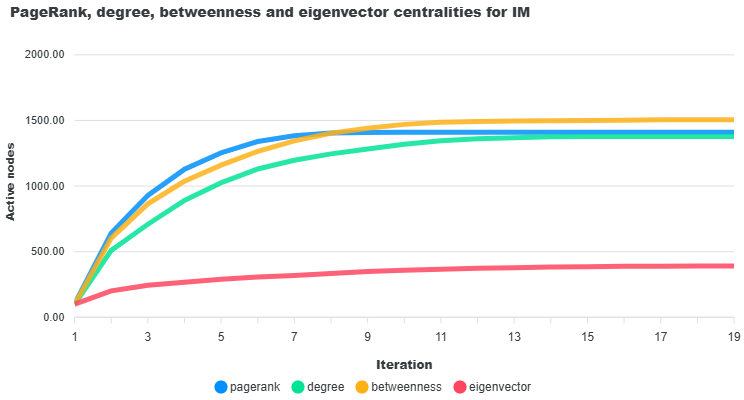}
        \label{fig:comparison_im}
    }
    \caption{IM: (a) Node counts (seed node selection with PageRank) (b) Comparison of total activation over time (seed node selection with PageRank, Degree, Betweenness, and Eigenvector centralities).}
\end{figure*}

In this study, we implement the IC model using an agent-based approach, where instead of looping over the edges, we iterate through each inactive agent while their neighbors try to influence them. For this, we set the influence probability, which denotes the success rate of a neighboring node activating the currently inactive node $v$, as $1/(in\_degree\_of\_v)$ \cite{ICModel}. An active node can try to influence its neighbors only once. \color{black}Initially, we select 100 active nodes with various centrality metrics. For our experiments, we use a Facebook dataset of social circles \cite{facebook_dataset}, which has 4039 nodes and 88,234 edges, and is available on the Stanford Large Network Dataset Collection \cite{snap_datasets}. 

To satisfy the condition that an active node should try to spread information only once, we utilize a temporary node type, ``Active Spreader," on top of the ``Active" and ``Inactive" states on which the model is based.  In this case, when a node is activated, it becomes an Active Spreader for one iteration, then changes its state to Active in the next iteration, where it cannot influence its neighbors anymore. The configuration to initialize the node types is given in Listing \ref{lst:infmax-nodetypes}.

\begin{lstlisting}[language=YAML, caption={Definition of node types in the IM simulation.}, label={lst:infmax-nodetypes} ]
nodetypes:
    Active_Spreader:
        choose_with_metric:
            metric: pagerank
            count: 100
    Active: 
        random-with-count:
            count: 0
    Inactive:
        random-with-count: 
            count: 3939
\end{lstlisting}

Following the network initialization, we store the influence probabilities for each edge as an edge parameter, calculated with the formula given in the previous paragraph. In each epoch, we iterate over the nodes. When an inactive node has an active spreader neighbor, we generate a number between 0 and 1. Each active spreader neighbor has the ability to activate the current node if the activation probability of the edge between them is larger than or equal to the current number. 

We run four experiments with different methods for the selection of initial active nodes: PageRank, degree, betweenness, and eigenvector centralities. Fig. \ref{fig:infmax-network} shows the initial state of the network and the activations that occurred after 20 iterations in the PageRank experiment. Fig. \ref{fig:page_rank_scatter} illustrates the number of nodes of each type in a scatter plot to demonstrate one of the available chart types in the tool.  While this data is collected by the framework in each iteration, researchers may need the total number of active nodes to put their results into formats such as tables and charts. For this purpose, we can write a simple data collection method that we will set to execute after each iteration.  

After conducting the PageRank experiment, we only change the selection metric in the configuration file, then re-run the simulation each time with the new setting. To display a comparison in a chart, we need to have a file containing the number of total active nodes in all four simulations, which we can do by using the merge methods provided by Crowd. The new file will show the values for each iteration labeled with the simulation name, which we utilize to draw Fig. \ref{fig:comparison_im}. Comparison figures such as Fig. \ref{fig:comparison_im} are commonly used in IM papers to show the improvement provided by the novel methods described. 

\begin{figure*}[h!]
    \centering
    \subfloat[]{
        \includegraphics[width=0.4\linewidth]{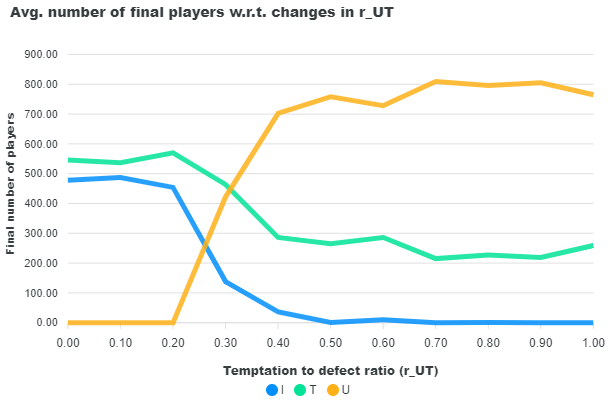}
        \label{fig:trust-game-charts1}
    }
    \hspace{0.04\linewidth}
    \subfloat[]{
        \includegraphics[width=0.4\linewidth]{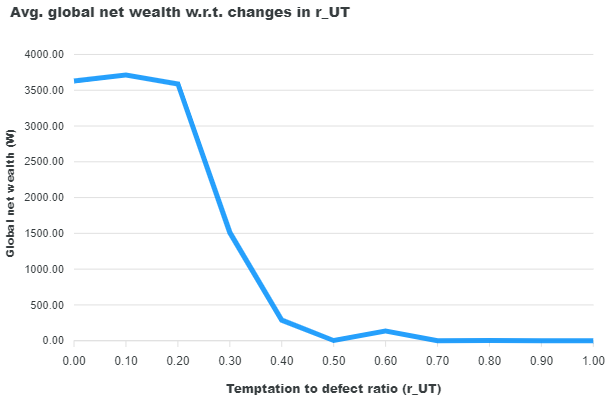}
        \label{fig:trust-game-charts2}
    }
    \caption{N-player trust game: (a) Averaged number of investors (I), trustworthy trustees (T), untrustworthy trustees (U). (b) Averaged global net wealth. Over 50 simulations for different temptation-to-defect ratios $r\_UT$ ($R\_T = 6$).}
    \label{fig:trust-game-charts}
\end{figure*}

In this implementation, we provide a basic example to illustrate how Crowd can be utilized to test the influence of selected seed nodes with various methodologies, combine the simulation results, and draw charts for comparison. With Crowd's simplified node initialization, data savers, merge methods, and built-in visualization, it provides an environment to run experiments with less implementation time. 

\subsection{Scenario 3: Trust Game on Networks}
\label{subsec:trust_game}
Cooperative games are commonly used to model the behavior of human society, where each player adopts a strategy that may result in varying benefits or costs.  Each type of game can introduce different kinds of roles and strategies. A common example is the Prisoner's Dilemma \cite{szabo}, which presents two strategies: being a cooperator who provides benefits to the other player by paying a cost, or a defector who denies benefiting another player, hence paying no cost and only benefiting from the cooperator's actions \cite{nowak}. The extension of this game to the populations can be inspected in structured and unstructured populations. In unstructured populations, the individuals can interact with any other individual with a similar possibility, while in structured populations such as networks, individuals interact with the people they are connected to, i.e., their neighbors.  The payoffs received by a player depend on the structure of the population.

An extension of cooperative games in economics, trust games, is based on investors and trustees. Investors have an initial fund and they transfer a portion of it to the trustee. In return, the trustee gives back an amount of their earnings, which can also be zero. In this section, we inspect an N-player evolutionary trust game in a network setting, a study by  Chica et. al \cite{nPlayerTrust}, and show how Crowd can be utilized for the simulation of evolutionary cooperative games. 

In this study, the agents represent the players, adopting one of the three strategies: Investor (I), trustworthy trustee (T), or untrustworthy trustee (U).  The agents can change strategies based on the proportional imitation rule, where the current agent's payoff is compared to that of a randomly selected neighbor. If the neighbor's payoff is higher, the agent switches to the neighbor's strategy with a probability that is set according to the expected improvement this change will provide. Similarly, the current payoff calculation for each agent is also done with respect to local factors, such as using a number of neighboring investors instead of global investors for the trustee's payoff formulas.  

In our implementation, we represent I, T, and U as node types, with strategy, current payoff, and previous payoff as node parameters. Global parameters, the received fund multipliers for trustworthy and untrustworthy trustees ($R\_T$ and $R\_U$, respectively), the temptation-to-defect ratio $r\_UT$, and minimum and maximum possible payoffs, are stored as network parameters. Constants are defined in the settings file, while parameters depending on network properties, such as the maximum possible payoff based on average degree, are set in the Python script before running the simulation. 

In our experiments, we use the Scale-Free (SF) network provided with the source code of the original study, which is generated using the Barabasi-Albert model with m = 3, and consists of 1024 nodes. While the original work experiments with various networks and agent strategies, we only analyze how the increase of the temptation-to-defect ratio $r\_UT$ affects the overall level of trust within the population. Following the settings provided in the original study, we increase $r\_UT$ from 0 to 1 in increments of 0.1 and run each configuration for 5000 iterations and 50 times. 

We utilize Crowd's \texttt{CustomSimNetwork} class, which allows the modelers to pass in the agent-level methods to be executed in each iteration. As mentioned in the previous paragraphs, agents adopt a proportional imitation strategy, comparing their payoffs with those of neighbors and changing their strategies probabilistically. The mathematical modeling of this strategy is further explained in  \cite{nPlayerTrust}. We implement proportional imitation using a custom method, which is applied as an agent method during each iteration of the simulation. 

\begin{figure*}[h!]
    \centering
    \subfloat[]{
        \includegraphics[width=0.33\linewidth]{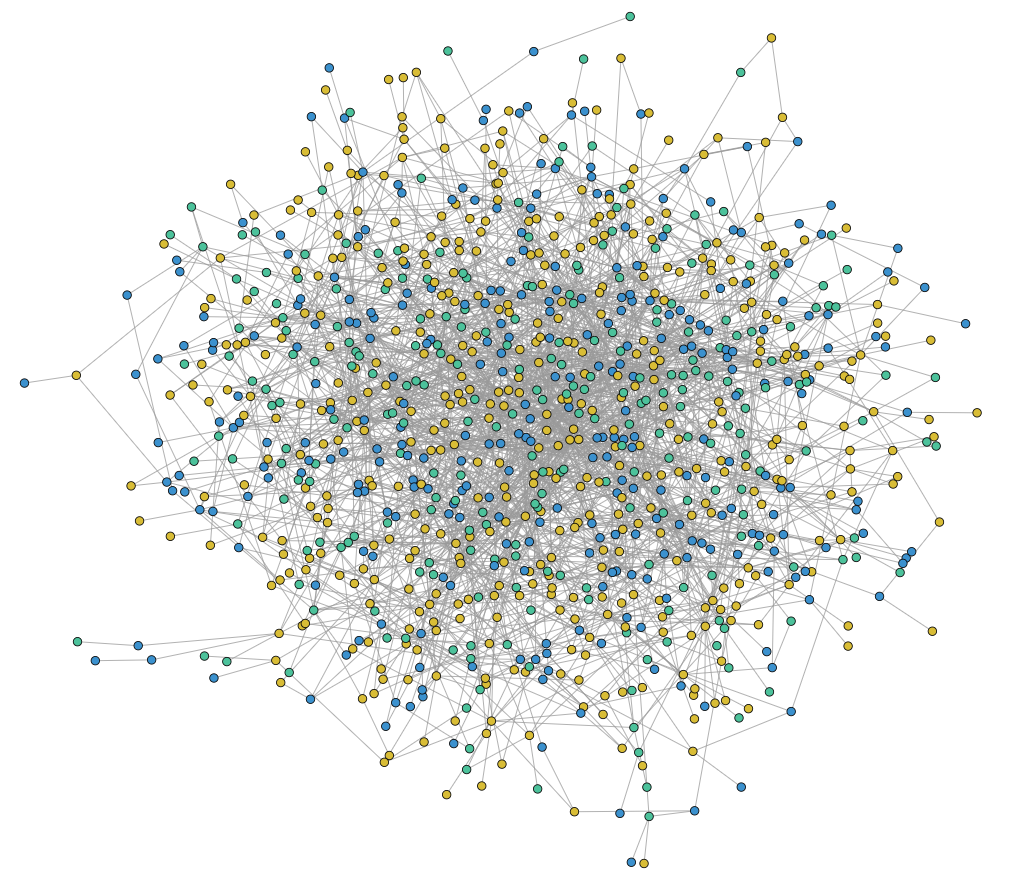}
        \label{fig:trust-game-initial}
    }
    \hspace{0.10\linewidth}
    \subfloat[]{
        \includegraphics[width=0.33\linewidth]{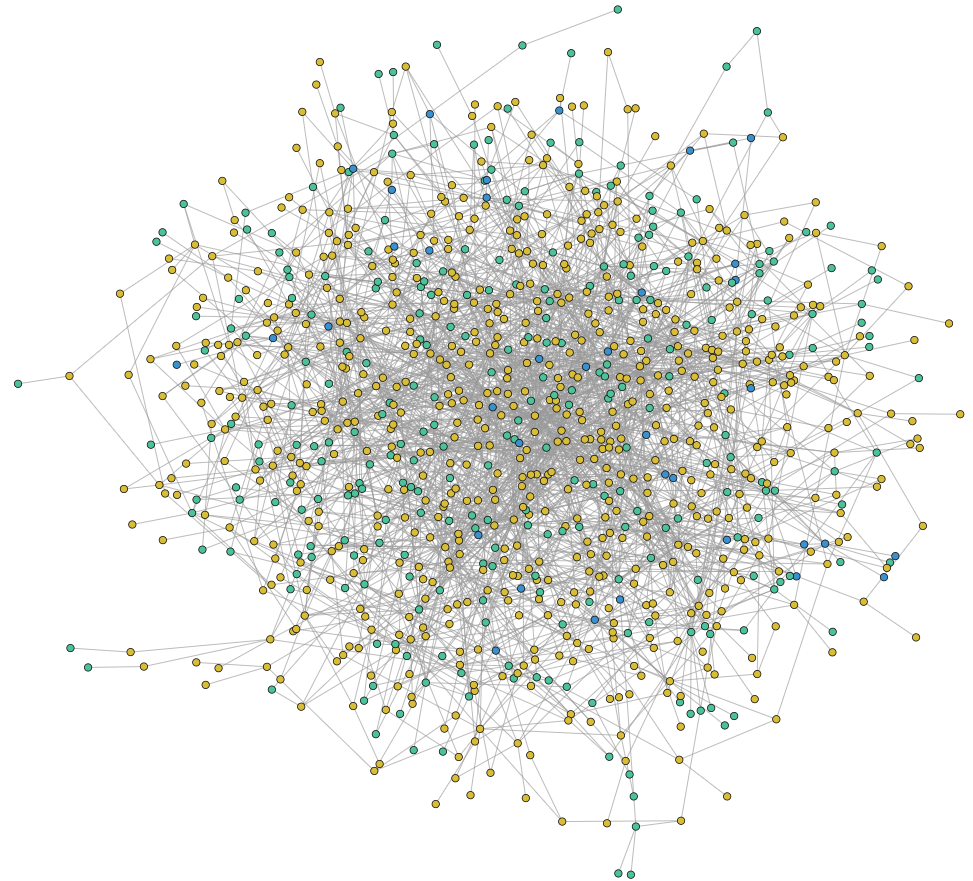}
        \label{fig:trust-game-final}
    }
    \caption{Network visualization of the trust game experiment: $n = 1000$ and $r\_UT = 0.4$ (blue: Investors (I), green: Trustworthy trustees (T), yellow: Untrustworthy trustees (U)). (a) Initial setting. (b) After 5000 iterations.}
    \label{fig:trust-game-network}
\end{figure*}

After executing this method for each agent, we then call the after-iteration method written for data collection and parameter-setting purposes. In this method, we calculate the new payoffs for each agent with its current strategy and sum up the individual payoffs to obtain a global payoff. We return the global payoff variable, which Crowd automatically stores its value to a file with the name of the method. Following the conclusion of each simulation execution, we run data collection methods to store the value of the $r\_UT$ variable and node type counts in the same file that is generated by the framework. 

After completion of the runs, we merge the results by taking the average of selected values with Crowd's ``merge in parent directory" feature. To visualize the effect of $r\_UT$, we further merge the results of simulations with varying $r\_UT$ values. With the data we obtained, we generate Figures \ref{fig:trust-game-charts1} and \ref{fig:trust-game-charts2} in Crowd's GUI. For Fig. \ref{fig:trust-game-charts1}, we utilize the numbers collected with the after-simulation methods. In this figure, we can see the number of investors and trustworthy trustees decreasing as more people adopt the untrustworthy strategy. Even though the agents switch to the untrustworthy type to gain more benefits, Fig. \ref{fig:trust-game-charts1} shows that the global net wealth declines as the number of untrustworthy agents increases. In addition to these charts, Fig. \ref{fig:trust-game-network} illustrates the network in the first and last iterations, highlighting the disappearance of investor agents and the takeover by untrustworthy agents across the network. Our results align with the findings of the original study, as Figures \ref{fig:trust-game-charts1} and \ref{fig:trust-game-charts2} show similar or nearly identical values, with slight differences likely due to the stochasticity of the system. Moreover, the dynamics observed in the simulation also follow the intertemporal sequential social dilemma definition, where short-term gains lead to choices that harm long-term collective outcomes \cite{multi_agent_survey}.

This case study provides an example of the use of Crowd's \texttt{CustomSimNetwork} class, showcasing how the framework can be utilized in a detailed analysis of cooperative games in structured populations represented as networks. Compared to other implementations of the scenario in different frameworks, Crowd enables quick modeling with minimal code. 

\section{Discussion}

\subsection{Modeling other social network problems with Crowd}
With the case studies explained in the previous section, we aimed to demonstrate how Crowd can be useful for three research areas in the field of social networks: epidemics, influence maximization, and trust games. In this section, we briefly address how Crowd can also be applied to other social network problems.

For instance, from a company's perspective to select influencers for product marketing \cite{influencerMarketing} or from a regular user's perspective to analyze methods to become an influencer \cite{becomeInfluencer}, engagements of posts on social media platforms can provide valuable insights. To model relationships such as followers, retweets, and replies in a network, one option is to utilize multiple networks, where each user will have the same node ID across all networks. Alternatively, a single network can be used with different types of edges, which can be represented using edge parameters in Crowd. For example, if person 1 retweets person 2’s tweets five times, the “retweet" parameter of the edge between them will be set to 5.

Besides numerical parameters, Crowd also supports categorical parameters. To represent relationships between users, an edge parameter such as “relationship” can be used, holding values like “colleague," “family," or “friend." These parameters—both numerical and categorical—can be used to model the level of influence in trend analysis or misinformation propagation studies.

Categorical node parameters can include information such as “language," “location," or “gender," which can be extracted from user profiles in online social networks. Numerical node parameters can be used to store values such as “age” (for epidemic simulations), “payoff” (for cooperative games), or “cluster ID" (after running community detection algorithms).

\subsection{Comparison with a general-purpose framework for the three case studies}
To demonstrate the use of Crowd in comparison with existing general-purpose frameworks, this section provides a direct comparison with \textit{Mesa} (v3.1.4) \cite{mesa} for the three case studies presented in this paper.  The comparison focuses on the required effort for modeling, visualization facilities, and execution times. Mesa is chosen as the baseline for comparison as it is the most widely used Python-based ABMS framework. Since Crowd is designed for network-based simulations, the comparison is limited to experiments that utilize network environments.

\subsubsection{Modeling stage}
Most ABMS frameworks require the user to implement Model and Agent classes. In the Model class, the network environment is initialized first, followed by setting up model parameters, creating agents, and specifying data collection settings. In Mesa, these steps must be explicitly defined in the code.  In Crowd, however, these configurations are specified in YAML files or selected via the GUI, eliminating the need for additional coding.

Moreover, in the Model class of Mesa, users create a \textit{run} method, which calls a self-defined \textit{step} function for a specified number of iterations. Within the \textit{step} function, both agent-level and model-level methods execute the simulation logic, followed by data collection. 

While this setup is not hard to implement, it is repetitive across simulations. Crowd automates this process by providing a \textit{run\_simulation} method within the \texttt{Project} class. Users can specify the number of iterations, snapshot periods, parameters for model exploration, and user-defined functions to execute at specific intervals. Unlike Mesa, data collection methods do not need to be explicitly passed to a data collector. If a method returns a value, Crowd automatically saves the results in a file. Further details on this process are discussed in Section \ref{subsec:network_architecture}, with an example in Section \ref{subsubsec:sim_dev_steps_def_methods}.

Additionally, Mesa does not provide methods to save the state of the graph or the collected data. Users must implement this manually to analyze the results of the simulation later or visualize the network using other tools. In Crowd, the graph is saved in the selected file formats at each snapshot period, and the collected data is saved at the end of the simulation.

\subsubsection{Execution times}
We conduct the experiments under the following conditions: Each case study is implemented in both frameworks as closely as possible, and each experiment is repeated 5 times. The average execution times are presented in the following tables. 

The first experiment was implemented on Google Colab and run on an L4 GPU. Due to the overhead introduced by LLM inference, a significant increase in the number of agents results in additional hours of runtime. Therefore, we conduct our experiments with 10, 50, and 100 agents. With each query to the LLM taking around 6 seconds on average, we complete a single experiment of 10 days within the times provided in Table \ref{tab:gabm_exec_times}. Since the number of nodes is relatively small, saving the network state in each iteration does not result in a significant overhead. Therefore, we save the network state in every iteration.

\begin{table}
    \centering
    \caption{Execution times of the generative agents case study in Mesa and Crowd frameworks}
    \begin{tabular}{|c|c|c|c|}
    \hline
        Tool/Nodes & 10 & 50 & 100\\ \hline
        Mesa & 10.58 min & 52.18 min & 107.07 min\\ \hline
        Crowd & \textbf{9.52 min} & \textbf{47.64 min} & \textbf{95.55 min} \\ \hline
    \end{tabular}
    \label{tab:gabm_exec_times}
\end{table}

For the second and third case studies, we save the graph data only in the first and last iterations, with the data collector's results being saved at the end of the simulation. These experiments are conducted on a system with an Intel i5-10210U CPU (1.60 GHz) and 32 GB RAM. 

For the second case study, influence maximization, we run our experiments for 20 iterations on three real-world networks from Facebook (4039 nodes, 88,234 edges), GitHub (37,700 nodes, 289,003 edges) \cite{githubDataset}, and Twitch (168,114 nodes, 6,797,557 edges) \cite{twitchDataset}. We demonstrate the change in the execution times as the number of nodes and edges increase in Table \ref{tab:inf_max_exec_times}.  It is important to note that the significant increase in the number of edges from the GitHub dataset to the Twitch dataset also contributes to the execution time.

\begin{table}
    \centering
    \caption{Execution times of the influence maximization case study in Mesa and Crowd frameworks on Facebook, GitHub, and Twitch social network datasets}
    \begin{tabular}{|c|c|c|c|}
    \hline
        Tool/Dataset & Facebook (4K)& GitHub (37K)& Twitch (168K) \\ \hline
        Mesa & \textbf{4.62 s} & 43.14 s &  16.22 min\\ \hline
        Crowd &  5.19 s & \textbf{26.27 s} &  \textbf{10.80 min} \\ \hline
    \end{tabular}

    \label{tab:inf_max_exec_times}
\end{table}

For the third case study, unlike in Section \ref{subsec:trust_game}, where the temptation to defect ratio is varied, here we run the experiments for 5000 iterations with data from the same study (Scale-Free networks generated with Barabasi-Albert model with m = 3) but with different network sizes. The execution times for experiments with 1024, 4900, and 14400 nodes are given in Table \ref{tab:trust_game_exec_times}. 

\begin{table}
    \caption{Execution times of the networked trust game case study in Mesa and Crowd framework}
    \centering
    \begin{tabular}{|c|c|c|c|}
    \hline
        Tool/Nodes & 1024 & 4900 & 14400\\ \hline
        Mesa &  38.03 s & 198.55 s & 11.42 min \\ \hline
        Crowd &  \textbf{24.24 s} & \textbf{117.91 s} & \textbf{7.53 min}\\ \hline
    \end{tabular}

    \label{tab:trust_game_exec_times}
\end{table}

As shown in Tables \ref{tab:gabm_exec_times}, \ref{tab:inf_max_exec_times}, and \ref{tab:trust_game_exec_times}, both frameworks perform similarly for these scenarios, while Crowd takes less time in most simulations. This illustrates that Crowd's additional simplifications do not slow down the execution and can be preferred for Python-based simulations without any performance drawback. 

\subsubsection{Visualization stage}
Mesa uses Python-based tools for visualization, allowing users to customize the visualization components through code. With this approach,  modelers have more flexibility in describing the appearance of graphs and charts. Crowd provides visualization primarily through its GUI, which does not require any coding. It utilizes JavaScript libraries and provides more interactive graphs and charts, allowing operations like zooming, dragging nodes and chart lines, and viewing node and link IDs.

\section{Conclusion and Future Work}

In this paper, we presented \textit{Crowd}, an open-source Python framework specifically developed to facilitate the modeling, simulation, and analysis of social networks. The network, agents, parameters, and various simulation settings are defined with a configuration file, allowing no-code diffusion modeling and a lower-entry barrier for social simulations. Further customization is enabled through user-defined Python methods that are executed in the time specified by the modeler. By focusing only on writing code for the specifics of their study and not requiring to handle the basics of simulation and visualization, researchers can quickly and simply address their research questions. 

In addition to a Python library, we provide a graphical user interface to create projects and simulations, visualize the network saved in each step of the simulation, draw charts, and aggregate results. Taking advantage of the Python ecosystem, the researchers can integrate any machine learning or data analysis libraries within their simulation and utilize novel methodologies, such as generative agents for more realistic modeling of human behavior compared to rule-based approaches.

In future work, we aim to expand Crowd's network generation capabilities with machine learning-based methods to synthesize data-driven network structures with user-specified requirements. Moreover, we plan to integrate an LLM-based code generator into the framework to assist both non-expert and experienced users with user-implemented agent and data collection methods. With these features, we intend to provide novel features to the agent-based and social network simulation spaces. 

In the current version of Crowd, only single-threaded execution of simulations is allowed. Utilizing Python's \textit{multiprocessing} module, we aim to add parallel processing to the framework and reduce the time needed for parameter sweeps by executing multiple simulations simultaneously. To further enhance Crowd's performance, we plan to experiment with compiling the library to C utilizing tools such as Nuitka. Additionally, we plan to add more scheduling options for custom network simulations and expand the library with additional visualization features to further extend the framework's modeling and analysis capabilities.

Furthermore, we plan to add facilities to Crowd to support the validation of simulation results, such as formatters for real-world data to be included in the list of data files to draw charts in the GUI, or simplified integration into post-simulation functions to perform desired computations with both data. We plan to test these functionalities with real-life datasets that provide information over time. 

Finally, we aim to establish a GitHub repository to share data and models that can be used within Crowd, supported by contributions from both our research group and the broader community. Feedback and participation in this repository will support the ongoing development and enhancement of Crowd.

% \section*{Acknowledgment}

\vspace{12pt}

\end{document}